\documentclass{ws-procs9x6}
\usepackage{amsfonts}
\newcommand{\be}{\begin{equation}}
\newcommand{\ee}{\end{equation}}
\newcommand{\bea}{\begin{eqnarray}}
\newcommand{\eea}{\end{eqnarray}}

\newcommand{\0}{\over }

\newcommand{\g}{g_{\rm eff}}
\newcommand{\geff}{g_{\rm eff}}

\newcommand{\im}{{\rm Im}\,}

\newcommand {\real}{\mathrm{Re}\,}
\newcommand {\imag}{\mathrm{Im}\,}

\newcommand {\cS}{\mathcal{S}}

\begin{document}

\title{
Anomalous specific heat in\\ ultradegenerate QED and QCD\\
}

\author{A. GERHOLD, A. IPP, A. REBHAN}

\address{Institut f\"ur Theoretische Physik, Technische
Universit\"at Wien, \\Wiedner Haupstr.~8-10, 
A-1040 Vienna, Austria}  

\maketitle

\abstracts{
We discuss the origin of the anomalous $T\ln T^{-1}$
behavior of the low-temperature
entropy and specific heat in ultradegenerate QED and
QCD and report on a recent calculation which is complete
to leading order in the coupling and which contains an infinite
series of anomalous terms involving also fractional powers
in $T$. This result involves dynamical hard-dense-loop resummation
and interpolates between Debye screening effects at larger
temperatures and non-Fermi-liquid behavior from only dynamically
screened magnetic fields at low temperature.
}

\section{Introduction}

Cold dense quark matter has important deviations from
Fermi-liquid behavior: below $T_c^{CSC} \sim 6$ to $60$ MeV
there is color superconductivity and long-range magnetic
interactions are responsible for the fact that the energy gap
is not proportional to $\exp(-c/g^2)$ as with short-range
interactions, but instead to $\exp(-c'/g)$.\cite{Son:1998uk,Rischke:2003mt}
Above $T_c^{CSC}$, and for unpaired quarks also below,
the only weakly (dynamically) screened magnetic
interactions are also responsible for an anomalous behavior
of entropy and specific heat, with a behavior
$C_v\sim\alpha_s N_gN_f \mu^2T\ln T^{-1}$
first discovered in the context of nonrelativistic QED by
Holstein, Norton, and Pincus\cite{Holstein:1973,Chakravarty:1995}.

In QED this effect is probably unobservably small (though it
may arise also from effective gauge field dynamics in systems of
strongly correlated electrons), but in QCD it can be orders
of magnitude larger because there are $N_g=8$ gauge bosons instead of
only one, and also $\alpha_s$ is much larger than $\alpha$.
However, more recently the existence of this effect had been
questioned by Boyanovsky and de Vega%
\cite{Boyanovsky:2000zj}, who instead found a $\alpha T^3 \ln T$
(when their renormalization-group summation of log's is undone).

In Ref.\cite{Ipp:2003cj}, we have recently confirmed the
correctness of the original result and succeeded in calculating
higher terms of the low-temperature expansion, for which only
the coefficient of the leading log was known before.
The higher terms involve cubic roots of temperature, which
can be traced to the fact that the frequency-dependent screening
length of quasi-static magnetic modes is
given by\cite{Weldon:1982aq}
$\kappa\simeq (\pi m_D^2\omega/4)^{1/3}$,
where $m_D$ is the (electric) Debye mass.

\section{Origin of the $T\ln T$ term}

The non-Fermi-liquid behavior is usually discussed in terms
of the spectral properties of the fermions.
The leading fermionic contribution to the entropy density can be written
as
\be
\cS_f\simeq
-4 N N_f \int {d^4K\over(2\pi)^4}
{\partial n_f(\omega)\over\partial T}
  \left(\imag\ln S_+^{-1}+\imag\Sigma_+\,\real S_+\right),
\ee
where it suffices to consider the particle $(+)$ as opposed to antiparticle
contribution.
The $T\ln T$ behavior can then be obtained\cite{Holstein:1973} 
from the singular behavior
of the fermion self-energy at the Fermi 
surface\cite{Holstein:1973,Brown:1999aq,Manuel:2000mk},
\be
\Sigma_+\simeq {g^2C_f\over24\pi^2}(\omega-\mu)
  \ln\left({M^2\over(\omega-\mu)^2}\right)+i{g^2C_f\over12\pi}|\omega-\mu|.
\ee

However, it is not a priori justified to leave out the contributions
from the gauge bosons. The entropy contributed by transverse modes
can be written to two-loop accuracy as
\be
\cS_T\simeq-2N_g\int{d^4K\over(2\pi)^4}{\partial n_b(\omega)\over\partial T}
\Bigl(\underbrace{\imag\ln D_T^{-1}}_{(A)}\underbrace{-\imag\Pi_T\,\real D_T}_{(B)}\Bigr).
\ee
The fact that quasi-static transverse gauge bosons are only weakly screened
by $\Pi_T\simeq-i{\pi m_D^2\omega/(4k)}$ leads to
nonanalytic behavior in $T$, which is exactly the same as that
of the interaction part of $S_f$: 
\be
S_f^{\rm int}\simeq S_{(A)}\simeq -{N_g m_D^2 T\036}\ln T^{-1}
\ee
It is only because $\cS_{(B)}\simeq-\cS_{(A)}$
that $S_f$ already gives the complete result to leading order.

If one organizes the calculation differently, as done
in Ref.\cite{Boyanovsky:2000zj} where the specific heat is
extracted from the internal energy, one in fact finds that
all $T\ln T$ terms as contributed by the fermions cancel out.
However, as we have shown in
Ref.\cite{Gerhold:2004tb}, the complete result is then coming
from the internal energy of the gauge boson sector, 
explicitly neglected in Ref.\cite{Boyanovsky:2000zj},
which
resolves the contradiction 
with the original results.

\section{Complete leading-order results}

It turns out that it is in fact advantageous to reorganize
the calculation such that all anomalous contributions come from
the gauge boson sector, by integrating out the fermionic
degrees of freedom first. This allows one to systematically calculate
beyond the leading-log approximation without having to calculate
the fermionic spectral densities beyond leading order. 
In Ref.\cite{Gerhold:2004tb} 
we have most recently shown that the infinite series of
anomalous contributions is contained in the following
hard-dense-loop (HDL)\cite{Braaten:1990mz} 
resummed expression valid for $T\ll\mu$:
\bea\label{Sfull}
&&{1\0N_g}(\mathcal S-\mathcal S^0)=-{\g^2\mu^2T\024\pi^2}
-{1\02\pi^3}\int_0^\infty dq_0\, {\6n_b(q_0)\0\6T}
\int_0^\infty dq\,q^2\,\biggl[ \nonumber\\&&\!\!
2\,\im \ln \left( q^{2}-q_{0}^{2}+\Pi^{\rm HDL} _{T} \0 q^{2}-q_{0}^{2} \right )
+\im \ln \left( \frac{q^{2}-q_{0}^{2}+\Pi^{\rm HDL} _{L}}{q^{2}-q_{0}^{2}}\right)
\biggr] + O(\g^4\mu^2T).\;\;\quad
\eea
Here $S^0$ is the ideal-gas value of the entropy density, and
$\g^2=g^2 N_f/2$ in QCD, but $e^2 N_f$ in QED.

At low temperature $T\ll\g\mu$, one finds that
the Landau damping cuts of the HDL propagators give rise to a series
of the form
\bea\label{Sseries}
{\mathcal S-\mathcal S_0\0N_g}&=& \frac{\geff ^{2}\mu^2 T}{36\pi ^{2}}
\left( \ln{4\g\mu\0\pi^2T}-2+\gamma_{E}-\frac{6}{\pi ^{2}}\zeta '(2)
\right)\nonumber\\&& 
+\, c_1 T^{5/3} + c_2 T^{7/3} +
c_3 T^3 (\ln(\g\mu/T)+c_4) + O(T^{11/3}),
\eea
where the coefficients up to and including $c_4$ can be found in
Ref.\cite{Gerhold:2004tb}.

\begin{figure}[t]
\centerline{\epsfxsize=3in\epsfbox{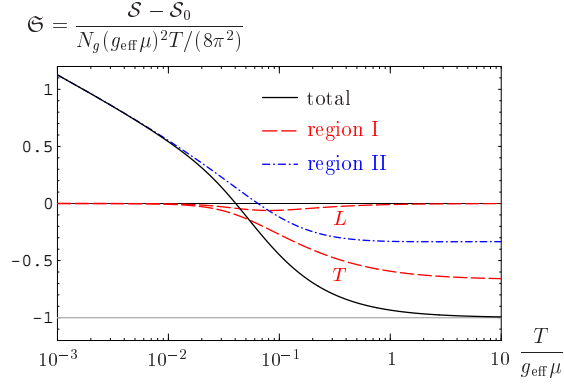}}
\caption{The function $\mathfrak S(T/(\g\mu))$
which determines the leading-order interaction contribution
to the low-temperature entropy.
The normalization is such that $\mathfrak S=-1$ corresponds
to the result of ordinary perturbation theory.
The dash-dotted line shows the contribution from spacelike momenta
(region II), comprising HDL Landau damping and
hard contributions; the two dashed lines
give the transverse (T)
and longitudinal (L) quasiparticle pole 
contributions (region I).
\label{figSdetails}}
\end{figure}

At $\g\mu \lesssim T \ll \mu$, one has also important quasiparticle
contributions, most of which are invisible in the low-$T$ expansion
since they involve terms suppressed by factors of $e^{-m_D/\sqrt3 T}$.
For $\g\mu \ll T \ll \mu$, one finally makes contact with more
familiar results from thermal perturbation theory, as $\mathcal S-\mathcal S_0
\to N_g(-\g^2\mu^2T^2/(12\pi^2)+\g^3\mu^3/(12\pi^4)+\ldots)$,
where it is the longitudinal
plasmon term $\propto m_D^3$ which makes (HDL) resummation
necessary. The expression (\ref{Sfull}) thus interpolates between
two physically rather different collective phenomena:
the plasmon effect from (electric) Debye screening, which only arises at
$T\gg \g\mu$, and non-Fermi-liquid effects from the only
dynamically screened magnetic interactions at $T\ll \g\mu$,
see Fig.~\ref{figSdetails}.

\section{Results for large and finite $N_f$}

The above result (\ref{Sfull}) is the leading-order result
for both QCD and QED at $T\ll\mu$. Higher-order terms either
involve extra powers of $g^2$ or $T^2/\mu^2$ (but not at the same time
higher powers of $\ln T$)\cite{Chakravarty:1995,Schafer:2004zf}.

\begin{figure}[t]
\centerline{\epsfxsize=3.2in\epsfbox{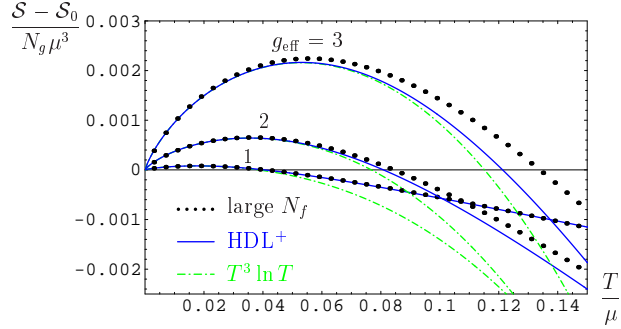}\qquad}
\caption{Complete entropy density
in the large-$N_f$ limit for the three values
$\g(\bar\mu_{\rm MS}=2\mu)=1,2,3$ (heavy dots), compared with
the full HDL result (solid line). Also given is the
low-temperature series up to and including the 
$T^3 \ln T$ contributions.\label{figSfull}}
\end{figure}

One case where we can actually investigate 
quantitatively the importance of
higher-order terms is in the exactly solvable limit of
large flavor number, which has been worked out for
finite chemical potential in Ref.~\cite{Ipp:2003jy}.
Figure \ref{figSfull} compares with the exact large-$N_f$ result
for $\g(\bar\mu_{\rm MS}=2\mu)=1,2,3$, and we find
that the HDL resummed result works well in the range where
the entropy exceeds its ideal-gas value. The exact large-$N_f$
result turns out to have even a slightly larger anomalous entropy.

\begin{figure}[ht]
\centerline{\epsfxsize=3in\epsfbox{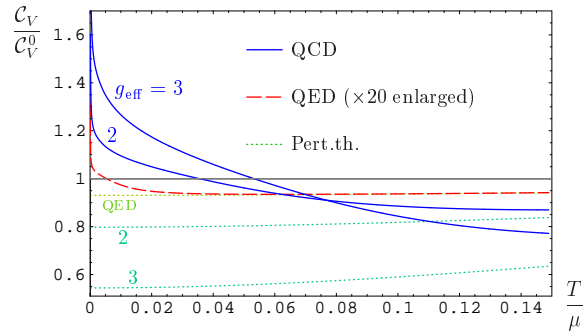}}
\caption{The HDL-resummed result for the specific heat
$\mathcal C_v$,
normalized to the ideal-gas value
for $\geff=2$ and 3 corresponding
to $\alpha_s\approx 0.32$ and $0.72$ in two-flavor QCD,
and
$\g\approx 0.303$ for QED.
The deviation of the QED result from
the ideal-gas value is enlarged by a factor of 20
to make it more visible.
\label{figspecificheat}}
\end{figure}

In Fig.~\ref{figspecificheat} we finally give results
for the specific heat at finite $N_f$, which shows its anomalous
behavior for a potentially interesting range in $T/\mu$.

\section*{Acknowledgments}
This work has
been supported by the Austrian Science Foundation FWF,
project no. 16387-N08 and by the \"OAD, project no. Amad\'ee-16/2003.


\end{document}